\documentclass[a4paper,11pt]{article}
\usepackage{pos}

\title{Detecting Lee-Yang/Fisher singularities by multi-point Padè}
\author*[a]{Francesco Di Renzo}
\author[b]{David Anthony Clarke}
\author[a]{Petros Dimopoulos}
\author[c]{Jishnu Goswami}
\author[d]{Christian Schmidt}
\author[d]{Simran Singh}
\author[e]{Kevin Zambello}

\affiliation[a]{DSMFI Università di Parma and I.N.F.N.,\\
  Parco Area delle Scienze 7/A, 43124 Parma, Italy}
\affiliation[b]{Department of Physics and Astronomy, University of Utah,\\
Salt Lake City, Utah 84112, United States}
\affiliation[c]{RIKEN Center for Computational Science, \\
Kobe 650-0047, Japan}
\affiliation[d]{Fakultät für Physik, Universität Bielefeld, \\
D-33615, Bielefeld, Germany}
\affiliation[e]{Dipartimento di Fisica, Università di Pisa and INFN,
  Sezione di Pisa, \\
Largo Pontecorvo 3, 56127 Pisa, Italy}

\emailAdd{francesco.direnzo@unipr.it}
\emailAdd{clarke.davida@gmail.com}
\emailAdd{petros.dimopoulos@unipr.it}
\emailAdd{jishnu.goswami@riken.jp}
\emailAdd{schmidt@physik.uni-bielefeld.de}
\emailAdd{ssingh@physik.uni-bielefeld.de}
\emailAdd{kevin.zambello@pi.infn.it}

\abstract{The Bielefeld Parma Collaboration has in recent years put forward a method to probe finite density QCD by the detection of Lee-Yang singularities. The location of the latter is obtained by multi-point Padè approximants, which are in turn calculated matching Taylor series results obtained from Monte Carlo computations at (a variety of values of) imaginary baryonic chemical potential. The method has been successfully applied to probe the Roberge Weiss phase transition and preliminary, interesting results are showing up in the vicinity of a possible QCD critical endpoint candidate.
In this talk we will be concerned with a couple of significant aspects
in view of a more powerful application of the method. First, we will
discuss the possibility of detecting finite size scaling of Lee-Yang/Fisher singularities in finite density (lattice) QCD. Second, we
will briefly mention our attempts at detecting both singularities in the complex chemical potential plane and singularities in the complex temperature plane. The former are obtained from rational approximations which are functions of the chemical potential at given values of the temperature; the latter are obtained from rational approximations which are functions of the temperature at given values of the chemical potential.}

\FullConference{The 40th International Symposium on Lattice Field Theory (Lattice 2023)\\
July 31st - August 4th, 2023\\
Fermi National Accelerator Laboratory\\}

\begin{document}
\maketitle

\section{Our workhorse: multi-point Padè}

Since a few years \cite{firstBiePr} the Bielefeld Parma collaboration has started a
project aiming to probe the QCD phase diagram by
reconstructing the singularity structure of the theory in the complex
chemical potential plane. The method is based on the approximation of
the relevant observables by a rational function via the so-called
multi-point Padè method.\\
The method works as follows. Suppose we know a few Taylor expansion coefficients of a given
function $f(z)$ at different points 
\begin{equation}
\ldots, \, f(z_k), \, f'(z_k), \ldots, f^{(s-1)}(z_k), \; \ldots \; \;
k=1 \ldots N 
\label{eq:OURdata}
\end{equation}
The basic idea of our multi-point Pad\'e approach is to approximate
(interpolate, actually) $f(z)$ by a convenient function. While a
polynomial approximation would be a natural choice with many respects,
that is not what we are interested in, because it would leave us
with no singularity pattern for $|z| \neq \infty$. Since we want
instead to guess the singularity structure of our $f(z)$, we 
consider the rational function $R^{m}_{n}(z)$
\begin{equation}
\label{eq:PadeRatFunct}
R^{m}_{n}(z) = \frac{P_m(z)}{\tilde{Q}_n(z)} = \frac{P_m(z)}{1+Q_n(z)} = \frac{\sum\limits_{i=0}^m \, a_i \, z^i}{1 + \sum\limits_{j=1}^n \, b_j \, z^j}\,
\end{equation}
with $m$ and $n$ being the degrees of the polynomials at numerator and denominator respectively.
We make a couple of preliminary observations which will be useful in
the following. First of all, 
writing $\tilde{Q}_n(z) = 1+Q_n(z)$ ensures that the rational function depends essentially on $n+m+1$ parameters. 
Having said that, we stress that {\em a priori}
we should naturally demand that there is no point $z_0$ such that
$ P_m(z_0) = \tilde{Q}_n(z_0) = 0$. The latter request seems indeed
natural: we should in principle exclude any (common) zero of both
numerator and denominator. Strictly speaking, if this were not the
case, 
we would have rather essentially defined the rational function
$R^{m'}_{n'}(z)$ with $n=n'+l$ and $m=m'+l$ for some integer $l>0$. 
We will nevertheless not exclude the possibility of common zeros, 
and we will instead live with that: as will see, the fact that common
zeros do show up will be a very frequent event.
We also make it clear that the number of coefficients we know can be different
at different points. For the sake of simplicity we will however assume
that $f^{(s-1)}$ is the highest order derivative which is known at
each point (together with all derivatives of degree $0 \leq g <
s-1$). \\
Now remember: we want $R^{m}_{n}(z)$ to be a good interpolation for
$f(z)$. It is quite obvious that we
want the rational function to account for all the information
available from (\ref{eq:OURdata}). In order for this to hold true, the
somehow simplest case is that of having $n+m+1=Ns$. If that is the
case, we will have
$$ \left(\frac{d}{dz}\right)^g R^{m}_{n}(z)|_{z=z_k} = f^{(g)}(z_k) $$
if we solve a system of equations, that is 
\begin{equation}
\label{eq:LinearProblem3}
\begin{split}
 & \vdots \\
P_m(z_k) - f(z_k)Q_n(z_k) &= f(z_k) \\ 
P_m'(z_k) - f'(z_k)Q_n(z_k) - f(z_k)Q_n'(z_k) &= f'(z_k) \\ 
 & \vdots \\
P_m^{(s-1)}(z_k) - f^{(s-1)}(z_k)Q_n(z_k) - \hdots &-f(z_k)Q_n^{(s-1)}(z_k)\\
&= f^{(s-1)}(z_k) \\
 & \vdots \\
\end{split}
\end{equation}
\noindent This is our recipe: by solving this system of linear
equations we determine the coefficients of the polynomials $P_m$ and
$Q_n$. In order to estimate the coefficients of the rational
functions we could of course rely on different methods, all somehow
related to the idea of minimizing a generalized $\chi^2$, {\em i.e.}
we could want to minimize the distance between the input Taylor 
coefficients and the relevant rational function, weighted by the 
errors available on the input coefficients (the latter will in the end
come from Monte Carlo measurements). Notice that this is
equivalent to solving an over-constrained system ($n+m+1<Ns$) in a least
squares sense. This has been compared to the linear solver method 
in \cite{BiePrPRD}.\\
Let us pause and inspect where we stand with respect to what we could
be interested in. A first observation is that the method leaves us
with an {\em interpolation} of $f(z)$, but of course we could be
interested in {\em extrapolating}, that is we could want to get some information on
$f(z)$ outside the interval where we collected the information encoded
in (\ref{eq:OURdata}). This is actually the case for lattice QCD. As a
matter of fact, this was for us a strong motivation for introducing the
method. This has to do with the (in)famous sign problem: for real values
of the baryonic chemical potential, lattice QCD computations by Monte
Carlo methods are hampered. The problem disappears for imaginary values,
and this has been largely relied on \cite{imagQCD1,imagQCD2}: one performs
computations where the latter are viable and then has to continue
results for real ({\em i.e.} physical) values of the chemical
potential. By our method we can compute a rational
approximation interpolating results obtained for imaginary values of
the baryonic chemical potential. The analytic continuation (that's
what we need) of the results is in our approach a most natural one:
simply compute our $R^{m}_{n}(\mu)$ for real $\mu$. Although
important, this is not the only thing we could be interested in. As
already pointed out, having a $R^{m}_{n}(z)$ as an interpolation 
for the function $f(z)$ enables us to guess the singularity structure of
$f(z)$: simply look at the poles of the rational function. Actually,
in the following this is what we are interested in.  

\section{The best case playground for the method: 2D Ising model}

In \cite{BiePrPRD} we applied for the first time the method
to lattice QCD. Namely, we were able to probe the Roberge Weiss
transition. By studying the number density at various temperatures and
for different imaginary values of $\hat{\mu}_B = \frac{\mu_B}{T}$
($\mu_B$ being the baryonic chemical potential), for a given value of
the temperature one should recognise
a phase transition taking place at $\hat{\mu}_B = i \pi$. 
By studying our rational approximants, we were able to inspect 
singularities at $\mbox{Im}\hat{\mu}_B = \pi$. At different values of
the temperature, the singularities take place closer and closer to the
imaginary axis as temperature gets closer and closer to $T_{RW}$, the
critical temperature of the Roberge Weiss transition. At each
temperature we registered $\mbox{Re}\hat{\mu}_{B 0}$, to be read as
the real part of the singularity with minimum real part. This quantity
was shown to verify the expected scaling in $T$ as the temperature was
approaching $T_{RW}$. An updated account on our study of the Roberge Weiss 
transition has been presented in \cite{CS_LAT23}. 
The singularities we have been talking about are
known as {\em Lee-Yang singularities}, simply related to zeros of the
partition function of the theory.\\
While the result in \cite{BiePrPRD} was a success, we can notice that
our result was obtained for a single scaling variable. Given our
setting, there are in
principle finite size errors in the procedure, which appeared to be
quite well under control. One would nevertheless like to apply the
method in a cleaner setting, in particular having the finite size
effects acting as main characters and not as minor ones. \\

This is what we did in \cite{IsingLAT22,Ising_arXiv}, to which we
refer the interested reader for more details: in the following we
will mainly sketch the conceptual path, to set the stage for the
experiments in lattice QCD. The theory which
we probed is the well-known Ising model in $D=2$
\begin{equation}
H = - J \sum_{<i,j>} \sigma_i \sigma_j - h \sum_i \sigma_i \;\;\;\;\;
(\sigma_i=\pm 1)
\end{equation}
We were able to probe both the {\em thermal singularities}, related to 
the so-called {\em Fisher zeros} and the {\em magnetic singularities}, 
related to the so-called {\em Lee-Yang zeros}. Roughly speaking,
Fisher zeros are values of $\beta$ at which the partition function at
$h=0$ is zero, while Lee-Yang zeros are values of $h$ at which the 
partition function at $\beta=\beta_c$ is zero, $\beta_c$ being
singled out by studying Fisher zeros. All this is associated to the 
phase transition taking place (at the critical temperature, 
{\em i.e.} at $\beta=\beta_c$) at $h=0$ 
and separating the paramagnetic phase from the ferromagnetic one. 
Here we recap what is going on with Lee-Yang zeros. For the case at
hand, the $f(z)$ of 
(\ref{eq:OURdata}) is the magnetization $m^{(L)}(h)$, which is
computed at various values of the lattice size $L$ as a function of the magnetic
field $h$, at $\beta=\beta_c$. 
The rational function in (\ref{eq:PadeRatFunct}) now reads 
$R^{m (L)}_{n}(h)$. All in all, we can recap
$$ f(z) \rightarrow m^{(L)}(h) \;\;\;\;\;\;\; 
R^{m}_{n}(z) \rightarrow R^{m (L)}_{n}(h)$$

\begin{figure}[ht]
	\centering
	\includegraphics{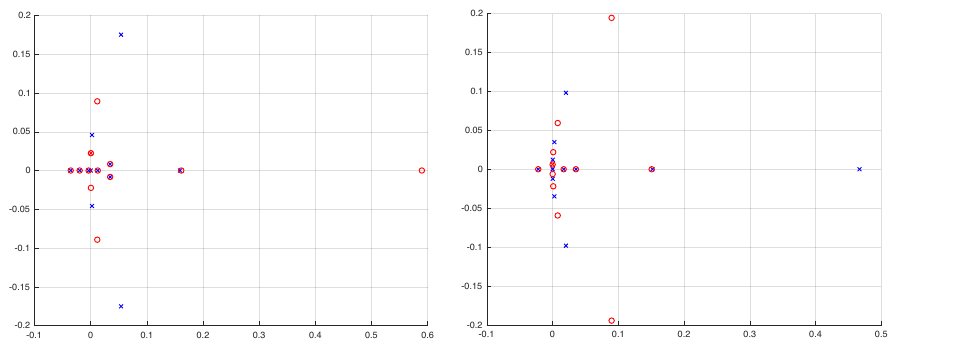}
	\caption{\label{fig:LeeYangZerosL}In the left panel we plot as
          blue crosses the zeros of the numerator of the rational
          approximant $R^{m (L)}_{n}(h)$ for the magnetisation on $L=15$;
          red circles are the zeros of the denominator. In the right
          panel we plot the same for $L=30$. In both cases we
          highlight the closest singularity to the real
          axis. Plots are in the complex $h$ plane.}
\end{figure}

\begin{figure}[hb] 
\centering
\includegraphics[scale=0.48]{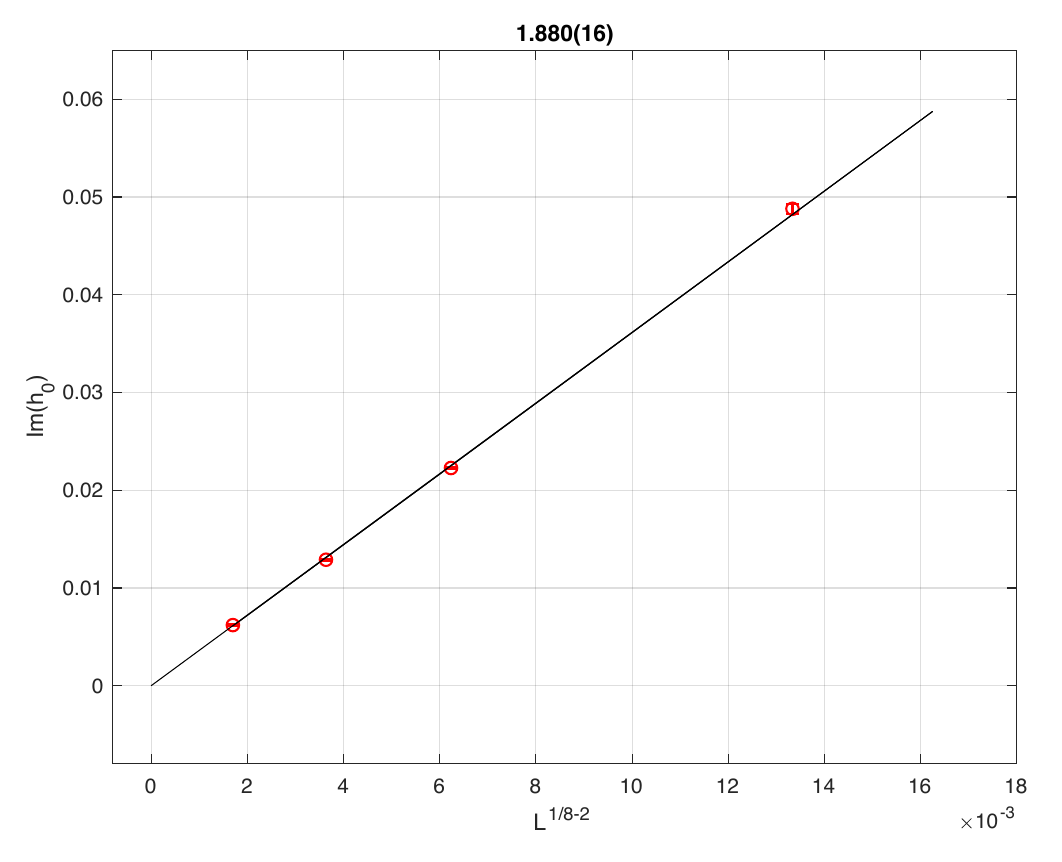}
\caption{\label{fig:ScalingLeeYangZerosL}As got from the rational
          approximants $R^{m (L)}_{n}(h)$, we plot the finite size
          scaling of the imaginary part of the singularity which is the
          closest to the real axis, $\mbox{Im}(h_0^{(L)})$. As expected, the
          phase transition appears to take place at $h=0$.}
\label{fig:LeeYangScaling}
\end{figure}

The dependence on the lattice size $L$ here is crucial. We first of
all compute the magnetization $m^{(L)}(h)$ at $\beta=\beta_c$ and 
various values of external magnetic field $h$ and lattice size
$L$. At each value of $L$, these results are interpolated by the $R^{m (L)}_{n}(h)$, which
display singularities which are the candidate Lee-Yang zeros
$h^{(L)}_0$, {\em i.e.} the 
singularity of the rational approximant for the  magnetisation which 
is the closest to the real axis. As a matter of fact, $h^{(L)}_0$ always sits
  at $\mbox{Re}(h^{(L)}_0)=0$, but with an imaginary part
  $\mbox{Im}(h^{(L)}_0)$ scaling in $L$ as 
\begin{equation}
\mbox{Im}(h^{(L)}_0) \sim L^{\cal{D}}
\end{equation}
$\cal{D}$ being a combination of critical exponents of the 2D Ising
model reading ${\cal{D}}=\frac{1}{8}-2$. 
Fig. \ref{fig:LeeYangZerosL} displays how our determinations of $\mbox{Im}(h^{(L)}_0)$
indeed turn out to get closer to the real axis as $L$ increases. 
Notice that a few zeros of the denominator are canceled by corresponding zeros of
  the numerator. These are not genuine pieces of information: actually
  their location vary if we vary {\em e.g.} the order of the Pad\'e
  approximant $[m,n]$. This is a consequence of numerical errors in our
  data. But the key point is that genuine pieces of information ({\em i.e.}
  actual zeros and poles) stay constant to a very good
  precision.
In Fig. \ref{fig:ScalingLeeYangZerosL} we depict the scaling of
$\mbox{Im}(h^{(L)}_0)$: to guide the eye, the horizontal axis is
$L^{\frac{1}{8}-2}$ ({\em i.e.} the theoretically expected power of
$L$), but the value we got ($1.88$) is indeed quite accurate. All in
all, the method is working beautifully. \\
Of course, we were not at all the first to study phase transitions by
Fisher/Lee-Yang zeros. For example, our method can be compared to
\cite{Finnish}: the good piece of information is that our method is
competitive.

\section{Can we do the same for Lattice QCD?}

The obvious question is: can we repeat the procedure for QCD? Now
that we have established a dictionary, what to do is pretty simple to
describe
\begin{equation}
m^{(L)}(h) \rightarrow 
\chi_1^{B (L)} (\hat{\mu}_B) = \frac{\partial}{\partial
  \hat{\mu}_B} \frac{\ln Z}{V T^3} \;\;\;\;\;\;\; \;\;\;\;\;\;\; 
R^{m (L)}_{n}(h) \rightarrow R^{m (L)}_{n}(\hat{\mu}_B) 
\end{equation}
where we have adopted for the number density the notation is terms of
its definition as a susceptibility ($\chi_1^{B}$). All the
measurements are here taken at $T=T_{RW}\sim 200 MeV$ on lattices
whose sizes are fixed by the spatial volume $(a N_s)^3$ 
(we adhere to the usual notation of
finite temperature QCD for the spatial size, {\em i.e.} $L=a N_s$, where $a$ is the lattice spacing).
We notice that our lattice regularization is a coarse one, given the
value $N_T=4$ for the (inverse) temperature in units of the lattice spacing. 
As for spatial sizes, these were dictated by our choice $N_s=12,16,20,24$. \\

\begin{figure}[hb]
	\centering
	\includegraphics[scale=0.4]{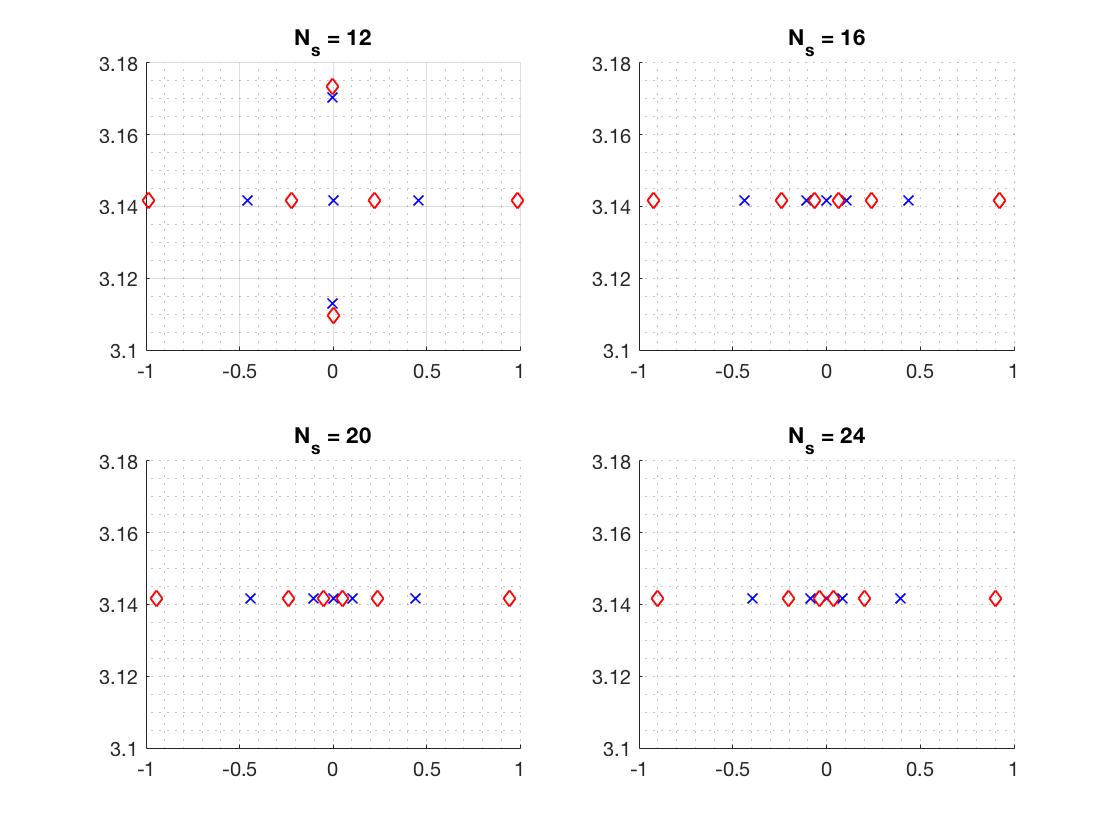}
	\caption{\label{fig:LeeYangZerosL_QCD}
 In the complex-$\hat{\mu}_B$ plane, we plot as
          blue crosses the zeros of the numerator of the rational
          approximant $R^{m (L)}_{n}(\hat{\mu}_B)$ for the number
          density of QCD computed at $T=T_{RW}\sim 200 MeV$;
          red circles are the zeros of the denominator. The various
          panels are for different values of $L=a N_s$. As expected, the
        singularities which are the closest to the imaginary axis get
        closer and closer to the expected Roberge Weiss transition
        point, {\em i.e.} $\hat{\mu}_B=i\pi$.}
\end{figure}

\begin{figure}[b] 
\centering
\includegraphics[scale=0.3]{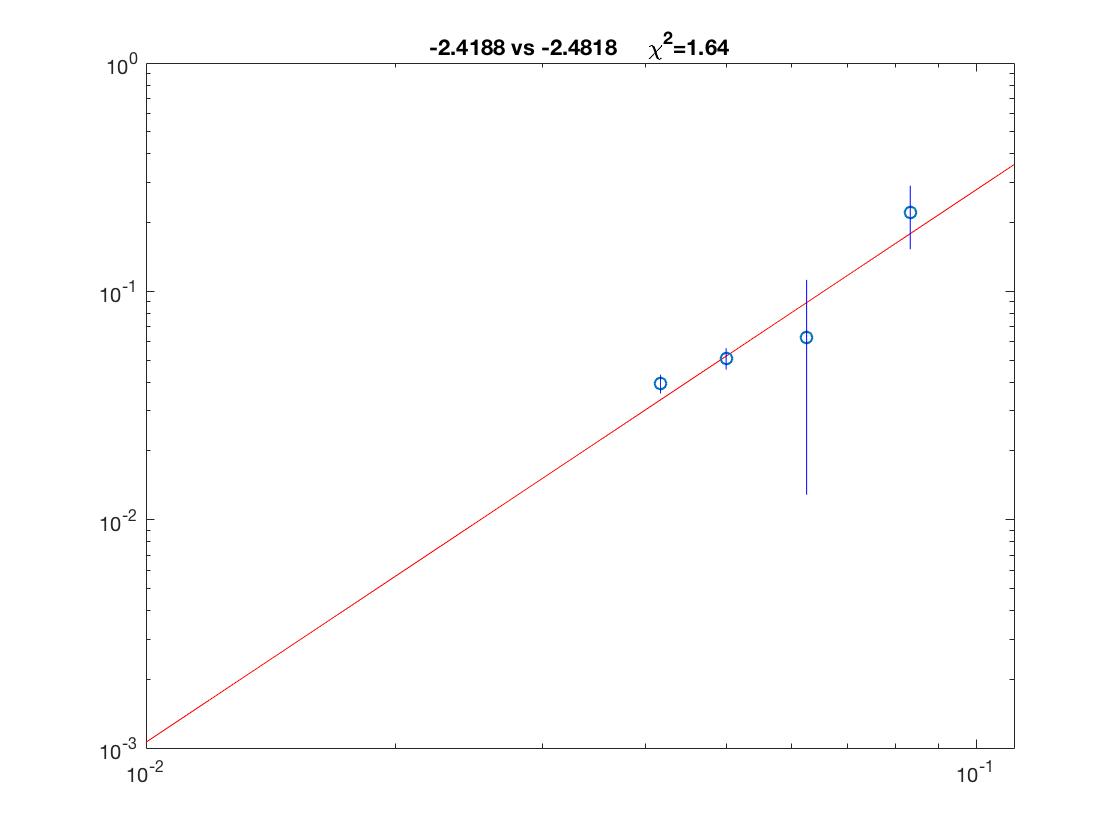}
\caption{\label{fig:LYzerosDifferentNs}For the rational
          approximants $R^{m (L)}_{n}(\hat{\mu}_B)$, we plot the real
          part of the singularity which is the
          closest to the imaginary axis, $\mbox{Re}(\hat{\mu}_{B
            0})$ (vertical axis) versus $L^{-1} \sim N_S^{-1}$ (horizontal
          axis). Notice how
          the error-bars are deformed in the log-log
          plot. The slope of the (finite size scaling) curve is very close to the expected
          one (but we are living with quite sizeable errors).}
\label{fig:LeeYangScaling}
\end{figure}

Fig.~\ref{fig:LeeYangZerosL_QCD} depicts the singularities which we
find for the rational approximation of our observable (the plot is in 
the complex-$\hat{\mu}_B$ plane). In particular, 
one can inspect the location of
$\mbox{Re}(\mu^{(L)}_{B 0})$, which is the real part of the 
singularity that at a given value of $L$ has real part 
that is the closest to zero. Notice that, as for imaginary parts, we
always find that they sit at $\pi$, that is to say that 
$\mbox{Im}(\mu^{(L)}_{B 0})=\pi$. As $L=a N_S$ gets larger and
larger, $\mbox{Re}(\mu^{(L)}_{B 0})$ gets closer and closer to zero:
this means that the method is doing a good job, at least at a
qualitative level.
To make all this quantitative, we have (as in the case of the Ising
model) to look at the scaling with the lattice size $L$. Also in this
case we should find a power law
\begin{equation}
\mbox{Re}(\mu^{(L)}_{B 0}) \sim L^{{\cal{D}}'}
\end{equation}
with ${\cal{D}}' = 2.4818\ldots$, once again fixed by a combination of
relevant critical exponents. As seen in 
Fig.~\ref{fig:LYzerosDifferentNs}, our analysis does not work that badly. The
power is very close to the exact one, with a reasonable value for the
$\chi^2$. Error-bars are quite important (and, by the way, they look
quite funny in the figure, due to the log-log plot), and thus we can
conclude that, while the picture is making sense, as in the case of
the Ising model, we need to refine the statistics and collect more
measurements in order to fully trust the procedure as effective.

\section{Outlook}

We reported on a first attempt at the study of finite size scaling in
the study of Lee-Yang zeros in QCD. This is not the only new
application of the multi-point Padè method we are developing. One can
actually study rational approximations of the form
$R^{m}_{n}(T;\hat{\mu}_B)$. Here the notation explicitly accounts for
the measurements being taken at different values of the temperature
$T$ for fixed values of $\hat{\mu}_B$. The approach is like that of
\cite{BiePrPRD}, {\em i.e.} in terms of a single scaling
variable, but this time singularities show up in the complex-T plane. 
While we presented very preliminary results at this year
conference, we are looking forward to reporting on a more precise
analysis in a not that long time. In a quite near future, we hope we
will also be able to confirm preliminary, interesting results showing 
up in the vicinity of a possible QCD critical endpoint candidate \cite{DavidLAT23}.

\section*{Acknowledgements}
This work has received funding from the European Union’s
Horizon 2020 research and innovation program under the Marie 
Skłodowska-Curie grant agreement No. 813942 (ITN EuroPLEx). FDR
acknowledges support by INFN under the research program 
{\em i.s. QCDLAT}.


\begin{thebibliography}{99}
\bibitem{firstBiePr}
C.~Schmidt, J.~Goswami, G.~Nicotra, F.~Ziesch\'e, P.~Dimopoulos, F.~Di
Renzo, S.~Singh and K.~Zambello,
{\em Net-baryon number fluctuations},
Acta Physica Polonica B, Proc. Suppl., 2021, \textbf{14}(2), 241,
doi:10.5506/APhysPolBSupp.14.241
\bibitem{BiePrPRD}
P.~Dimopoulos, L.~Dini, F.~Di Renzo, J.~Goswami, G.~Nicotra, C.~Schmidt, S.~Singh, K.~Zambello and F.~Ziesch\'e,
{\em Contribution to understanding the phase structure of strong
  interaction matter: Lee-Yang edge singularities from lattice QCD},
Phys. Rev. D \textbf{105}, no.3, 034513 (2022)
doi:10.1103/PhysRevD.105.034513
\bibitem{imagQCD1}
P.~de Forcrand and O.~Philipsen,
%``{\em The QCD phase diagram for small densities from imaginary chemical potential},
Nucl. Phys. B \textbf{642}, 290-306 (2002),
doi:10.1016/S0550-3213(02)00626-0;
\bibitem{imagQCD2}
M.~D'Elia and M.~P.~Lombardo,
{\em Finite density QCD via imaginary chemical potential},
Phys. Rev. D \textbf{67}, 014505 (2003),
doi:10.1103/PhysRevD.67.014505
\bibitem{CS_LAT23}
C.~Schmidt, D.A.~Clarke, J.~Goswami, P.~Dimopoulos, F.~Di
Renzo, S.~Singh, V.~V.~Skokov and K.~Zambello,
PoS \textbf{LATTICE2023}, 167 (2024)
\bibitem{IsingLAT22}
F.~Di Renzo and S.~Singh,
{\em Multi-point Pad\`e for the study of phase transitions: from the Ising model to lattice QCD},
PoS \textbf{LATTICE2022}, 148 (2023),
doi:10.22323/1.430.0148
\bibitem{Ising_arXiv}
S.~Singh, M.~Cipressi and F.~Di Renzo,
{\em Exploring Lee-Yang and Fisher Zeros in the 2D Ising Model through Multi-Point Pad\'e Approximants},
arXiv:2312.03178 [hep-lat]
\bibitem{Finnish}
A.~Deger and C.~Flindt,
{\em Determination of universal critical exponents using Lee-Yang theory},
Phys. Rev. Research. \textbf{1}, 023004 (2019),
doi:10.1103/PhysRevResearch.1.023004
\bibitem{DavidLAT23}
D.A.~Clarke, J.~Goswami, P.~Dimopoulos, F.~Di
Renzo, C.~Schmidt, S.~Singh and K.~Zambello,
PoS \textbf{LATTICE2023}, 168 (2024)
\end{thebibliography}
\end{document}